\documentclass[11pt,a4paper]{article}
\pdfoutput=1
\usepackage{jheppub}

\usepackage{amsmath}
\usepackage{verbatim}
\usepackage{graphicx}
\usepackage{mathrsfs}
\usepackage{appendix}
\usepackage{caption}
\usepackage{float}
\usepackage{subfig}
\usepackage{mathtools}
\usepackage{tabularx}
\usepackage{bigints}
\usepackage{wasysym}
\usepackage[normalem]{ulem}

\newcommand{\e}{\epsilon}

\newcommand{\be}[1]{ \begin{equation}\label{#1} }
\newcommand{\ee}{\end{equation}}
\newcommand{\bea}[1]{\begin{eqnarray}\label{#1} }
\newcommand{\eea}{\end{eqnarray}}
\newcommand{\bes}{\begin{subequations}}
\newcommand{\ees}{\end{subequations}}

\newcommand{\p}{\partial}
\renewcommand{\a}{\alpha}
\renewcommand{\b}{\beta}
\renewcommand{\t}{\tau}

\newcommand{\refb}[1]{(\ref{#1})}

\renewcommand{\(}{\left(}
\renewcommand{\)}{\right)}

\newcommand{\non}{\nonumber}
\newcommand{\lb}{\left[}
\newcommand{\rb}{\right]}

\title{Galilean Electrodynamics: Covariant formulation and Lagrangian}

\author[a]{Aditya Mehra,} \author[b,c]{Yaman Sanghavi.} \author{\\}
 \affiliation[a]{Max-Planck-Institut f{\"u}r Gravitationsphysik, Albert-Einstein-Institut, 14476, Golm, Germany.\\} 
\affiliation[b]{Stony Brook University, New York, 11794, USA.\\}
\affiliation[c]{Indian Institute of Technology Kanpur, Kanpur 208016, INDIA.}
\emailAdd{aditya.mehra@aei.mpg.de}
\emailAdd{yaman.sanghavi@stonybrook.edu}

\abstract{In this paper, we construct a single Lagrangian for both limits of Galilean electrodynamics. The framework relies on a covariant formalism used in describing Newton-Cartan geometry. We write down the Galilean conformal algebra and its representation in this formalism. We also show that the Lagrangian is invariant under the Galilean conformal algebra in $d = 4$ and calculate the energy-momentum tensor.
}

\preprint{}

\begin{document}

\maketitle

\newpage

\section{Introduction}
Relativistic classical electrodynamics is a theory that studies the interactions between currents and electric charges. In the classical regime where the quantum mechanical effects are negligible, this theory describes electromagnetic phenomena. However, for small distances, these interactions are best expressed by quantum electrodynamics.

In the development of covariant field theories of radiation and matter, this theory provides both a point of reference and a point of departure.
The observation predicted by Maxwell's equations about a universal speed of light in vacuum gave birth to the special theory of relativity.
The symmetries of special relativity made us realise that the equations of classical electrodynamics can be written down in a compact form, using a Lorentz vector representing the field, where we can maximally simplify various dynamical properties in a way that the Lorentz covariance is easily manifest.

Classical electrodynamics is also an example of a conformal field theory (CFT). It is Poincar\'e invariant in all dimensions but conformally invariant in $d=4$ \cite{Jackiw:2011vz}. Invariance under finite-dimensional conformal transformations in the covariant formulation has been discussed in \cite{Jackiw:2011vz, Bagchi:2014ysa}. This theory is anomalous at the quantum levels, but when we generalise to $\mathcal{N}=4$ Supersymmetric Yang-Mills (SYM), the conformal symmetries survive miraculously \cite{Beisert:2010jr, Beisert:2017pnr, Beisert:2018zxs}. One of the reasons we study conformal invariance in gauge theories is the AdS/CFT correspondence \cite{Maldacena:1997re} which provides a concrete example of the Holographic Principle \cite{tHooft:1993dmi, Susskind:1994vu}. The Holographic Principle tells us about the connection between a theory of gravity in $d+1$ dimensions to a quantum field theory in $d$ dimensions. The AdS/CFT correspondence perceives it by relating a specific theory of gravity (Type IIB superstring theory) on 5-dimensional negatively curved Anti-de sitter times $S^5$ to a specific quantum field theory ($\mathcal{N} = 4$ $SU(N)$ Supersymmetric Yang-Mills theory) in $d = 4$.

One thing to be noted is that the conformal symmetry is infinite-dimensional in two dimensions \cite{Belavin:1984vu}. It means one can compute various physical quantities without even resorting to a Lagrangian description.
The infinite symmetries apply only to field theories found in $d=2$. But for classical electrodynamics that is conformally invariant in $d>2$, it is not applicable. It would be interesting to have infinite symmetries in all possible spacetime dimensions (especially in $d=4$ for our case) such that it will be of use to constrain various quantities. If one looks at the non-relativistic limits of conformal field theories, this wish gets fulfilled. The conformal algebra in non-relativistic limit gives rise to infinite-dimensional Galilean conformal algebra (GCA) \cite{Bagchi:2009my}.

Galilean electrodynamics (GED) is one such example of a Galilean conformal field theory that exhibits infinite-dimensional Galilean conformal symmetries in four spacetime dimensions. Galilean Electromagnetism was first examined in the language of electric ($\vec{E}$) and magnetic ($\vec{B}$) vectors by Le Ballac and Levy-Leblond in 1973 \cite{LBLL}. In \cite{Bagchi:2014ysa}, the authors reformulate GED by taking the non-relativistic scaling on spacetime coordinates as well as on  electromagnetic four-potential vector $(A_{\mu})$. They started with the equations of motion of the original relativistic theory and expressed it in terms of the component form of Lorentz vector $(f_{\mu}= f_t, f_i)$. Then they applied the non-relativistic limit to get the two limits (Electric and Magnetic limit) of GED. The possible disadvantage of this process of taking the limit on CFTs is that there is no standard way to do the same on the Lagrangian. But now, there has been a lot of progress to find the action of GED in recent years by using different techniques in hand \cite{Duval:2014uoa, Festuccia:2016caf, Bergshoeff:2015sic, Bleeken:2015ykr, Banerjee:2019axy, Hansen:2020pqs}. In \cite{Chapman:2020vtn}, the authors went one step forward where they couple this theory to a Schr\"odinger scalar in 2+1 dimensions to study the quantum properties. They found an infinite number of couplings at the quantum level due to the scalar field present in the Galilean multiplet of the gauge field.

The main reason for looking into GED is that the conformal symmetry survives at the quantum regime of relativistic $\mathcal{N} = 4$ SYM. We expect that the non-relativistic $\mathcal{N} = 4$ SYM will have invariance under GCA both at classical and quantum domains. The presence of such a non-relativistic sector raises fascinating possibilities due to the existence of infinite-dimensional symmetries. If this is a close sub-sector, it could turn out to be a new integrable sub-sector like the planar limit of $\mathcal{N} = 4$ SYM.

As we notice, all the techniques rely mainly on the component formalism of the Lorentz vector to proceed. It motivates us to frame a covariant formulation for GCA and Galilean electrodynamics in this paper. We first start with some features of the geometry of Galilean space-time which are pertinent to our formalism of GCA and Galilean electrodynamics. We then give a brief review of Galilean conformal algebra in component form. Once we have them, we write GCA and its representation in covariant formalism. We can write a single Lagrangian that defines the dynamics of both the limits of GED in four dimensions which was not possible before. We also look at the invariance of the Lagrangian under GCA. Finally, we calculate the energy-momentum tensor and find the conserved Noether currents associated with this theory.

\section{Galilean Conformal Algebra}
\subsection*{Geometry of Galilean Spacetime: Newton-Cartan geometry}\label{sdll}
We will first discuss the geometry of Galilean spacetime and then move to Galilean conformal algebra. We know from literature \cite{LBLL} that Lorentz transformations of a vector $V_r^\mu = (V_r^0, V_r^i)$ for boost velocities $v <<c$ in four dimensions are given by
\bea{1}
V_r'^0 = V_r^0-\frac{v^i}{c} V_r^i + O\(\frac{v}{c} \)^2,~~V_r'^i = V_r^i-\frac{v^i}{c}V_r^0 + O\(\frac{v}{c} \)^2.
\eea
where $i=(1,2,3)$, $c$ is the speed of light and $r$ denotes the relativistic quantities. If we take $c \to \infty$ directly in \refb{1}, it won't lead to Galilean transformations. For that we have to first classify vectors into two categories:
\begin{itemize}\label{dc}
\item \textbf{Case 1}: The vectors $V_r^\mu$ for which the ratio ($V_r^0/V_r^i$) is proportional to $`c$' in the limit $c \to \infty$. Such vectors are often called largely timelike vectors and they can be written as
$V^\mu_r = (cV^0 , V^i)$.
By substituting it in \refb{1}, we see that the $V^\mu = (V^0, V^i)$ transforms like a Galilean vector in the limit.
One such example is the spacetime coordinates: 
$x_r^\mu = (ct , x^i)$ $\implies$ $x^\mu = (t , x^i)$.
\item \textbf{Case 2}: The vectors $V_r^\mu$ for which the ratio ($V_r^0/V_r^i$) is inversely proportional to $`c$'. Such vectors are known as largely spacelike vectors. They can be mathematically realized as
$V^\mu_r = \(-V^0/c , V^i\)$. The vector $V_\mu = (V_0, V_i)$ also transforms like a Galilean vector in the limit $c \to \infty$.
The examples include 4-force:
${\mathcal{F}_r}_\mu = (F.v/c , F^i)$ $\implies$ $\mathcal{F}_\mu = (-F.v , F^i)$
\end{itemize}
These reparametrized vectors $V^\mu$ and $V_\mu$ will act as the Galilean versions of Lorentz vectors in the non-relativistic limits. One can verify that these vectors transform under Galilean boost and rotations in accordance with Galilean relativity.
We can write down how these vectors change under general coordinate transformations $x^\mu \to x'^\mu(x^\nu)$:
\be{10a}
V^\mu \to V'^\mu = \frac{\p x'^\mu}{\p x^\nu} V^\nu,~V_\mu \to V'_\mu = \frac{\p x^\nu}{\p x'^\mu} V_\nu.
\ee
Therefore, we call $V^\mu$ as a contravariant vector and  $V_\mu$ a covariant vector. For generalisation to Galilean tensors, we urge the readers to look at Appendix[\ref{GTs}].

We have two kinds of vectors \eqref{10a}, those represented by a contravariant vector and the others which are represented by a covariant vector. Therefore, we can have two kinds of metric tensors, represented by $\t_{\mu\nu}$ and $h^{\mu\nu}$ respectively for contravariant and covariant vectors. These metric tensors have been known in the literature \cite{Kuenzle:1976sdk, Dautcourt:1990sds, Andringa:2010it, Bleeken:2015ykr} and are given by
\bea{14}
h^{\mu\nu} = \begin{bmatrix*}[r]
0 & \phantom{-}0 \\
0 & \phantom{-}1_{3\times 3}\\
\end{bmatrix*},
\quad
\tau_{\mu\nu} = \begin{bmatrix*}[r]
-1 & \phantom{-}0\\
0 & \phantom{-}0_{3\times 3}\\
\end{bmatrix*}.
\eea
The metric $h^{\mu\nu}$ preserves the spatial norm of an up-indexed vector under rotations and Galilean boosts whereas, $\tau_{\mu\nu}$ preserves the 0-component of a down indexed vector under rotations and Galilean boosts. Note that these metric tensors are non-invertible. Moreover, $\tau_{\mu\nu}$ has a matrix rank equal to 1, therefore it can be written as
\bea{15}
\tau_{\mu\nu}=-\tau_\mu \tau_\nu ~~\text{where}~~\tau_\mu =\begin{bmatrix*}[r]
1 & 0 & 0 & 0
\end{bmatrix*} .
\eea
The following relation holds between the metric tensors that will be used extensively later in the paper
\be{17}
h^{\mu\nu}\tau_{\nu}=0.
\ee
In addition to these metric tensors, there are two more tensors defined in the literature. These two will act like the metric tensors for the irreducible subspaces of the contravariant (up-indexed vectors) vector space and the covariant (down-indexed vectors) vector space under the action of Galilean boosts and rotations. The irreducible subspace of the contravariant vector space is a 3-dimensional vector space defined by
\be{T0}
T_0 := \{V^\mu \; | \; V^\mu \t_\mu = 0 \}.
\ee
Similarly, the irreducible subspace of the covariant vector space is a 1-dimensional vector space defined by
\bea{C0}
C_0:= \{V_\mu \; | \; V_\mu h^{\mu\nu} = 0 \}.
\eea
It is easy to verify that these two vector spaces defined above are invariant and irreducible under the action of Galilean boosts and rotations. Therefore they deserve their own metric tensors. We will call them submetric tensors and denote them by $h_{\mu\nu}$ for $T_0$ and $\t^{\mu\nu}$ for $C_0$. The submetric $h_{\mu\nu}$ turns out to be 
\be{}
h_{\mu\nu} = \begin{bmatrix*}[r]
a & \phantom{-}b_i \\
b_i & 1_{3\times 3}\\
\end{bmatrix*},
\ee
where $a,b_i (=b_1,b_2,b_3$) can be any real numbers. Under Galilean boosts and rotation, the $3\times 3$ identity matrix block doesn't change but the rest of the components $a,b_i$ can change.
The submetric $\t^{\mu\nu}$ turns out to be
\be{}
\t^{\mu\nu} = \t^{\mu} \t^{\nu} \;\; \text{where} \; \t^\mu = (1, c_1, c_2, c_3),
\ee
where $c_i$'s can be any real numbers and they will change under boosts and rotations.
\subsection*{Properties of the metric and the submetric tensors}
We know that the submetric tensors have some components which transform under Galilean boosts and rotations. We cannot use such components to calculate physical scalars. Still, the submetric tensors can be combined with the metric tensors to create quantities that do not change under boosts and rotations. One such useful property to note about the submetric $\t^\mu$ is $\t^\mu \t_\mu = 1$ whereas another property for $h_{\mu\nu}$ is given by $h^{\mu\a}h_{\a\b} h^{\b\nu} = h^{\mu\nu}$.
In addition, we can construct a new and very useful tensor out of these. To find that, we first notice a curious combination of $\t_\mu$ and some arbitrary down indexed vector $V_\mu$ defined as
\be{}\label{hgf}
V^\star_{\mu\nu}:= 2 \t_{[\mu} V_{\nu]} =(\t_\mu V_\nu - \t_\nu V_\mu).
\ee
The only non-zero components of $V^\star_{\mu\nu}$ are $V^\star_{0 i} = - V^\star_{i0} = V_i$. This antisymmetric combination doesn't have the $V_0$ component at all. Now, notice that the only components of $h_{\mu\nu}$ that are invariant under Galilean boosts and rotations are $h_{ij}$ where $i$ and $j$ run from $1$ to $3$. Therefore, we can use an antisymmetric combination of $\t_\mu$ with $h_{\a\b}$ so that the non-zero components of the resulting combination will only have the information of $h_{ij}$ in it. This means that the resulting tensor will be invariant under Galilean boosts and rotations. That tensor is defined as
\bea{T}
T_{\a\b\mu\nu} := 4 \t_{[\a} h_{\b] [\mu} \t_{\nu]} = (\t_\a h_{\b\mu} \t_\nu - \t_\b h_{\a\mu} \t_\nu - \t_\a h_{\b\nu} \t_\mu + \t_\b h_{\a\nu} \t_\mu).
\eea
It is easy to see that $T_{\a\b\mu\nu}$ remains same on exchanging $\a$ and $\nu$  simultaneously with an exchange of $\b$ and $\mu$ that is given as
$T_{\a\b\mu\nu} = T_{\nu\mu\b\a}$.
Notice that $T_{\a\b\mu\nu}$ changes sign if we exchange $\a$ with $\b$ or $\mu$ with $\nu$ i.e.
\be{}
T_{\a\b\mu\nu} = - T_{\b\a\mu\nu} = - T_{\a\b\nu\mu} = T_{\b\a\nu\mu}.
\ee
It also follows the property given by
\be{}
T_{\a\b\mu\nu} - T_{\a\mu\b\nu} = T_{\a\nu\mu\b}.
\ee
The advantage of this tensor $T_{\a\b\mu\nu}$ is that it encompasses the geometrical properties of the metric tensors and the submetric tensor $h_{\mu\nu}$ in such a way that it remains unchanged under rotation and Galilean boosts. We will use it in the construction of Galilean conformal algebra in covariant formulation in the next section.

For completeness, we discuss the partial derivative $\p_\mu$ with respect to space-time coordinates and also construct a contravariant version of this. We know from the chain rule of derivatives, $\p_\mu := (\p_t,\p_x,\p_y,\p_z)$ follows the same transformation that of a covariant Galilean vector. Its transformation is given by
\be{p1}
\p'_\mu = \frac{\p x^\nu}{\p x'^\mu} \p_\nu.
\ee
Unlike the relativistic case, the metric tensors are non-invertible here which means that there is no reversible way to convert an up index to down index or vice versa. Nevertheless, given a vector $V^\mu$, we can create a down indexed vector $\tilde{V}_\mu = V^\nu \tau_{\mu\nu}$. If we have only the information of $\tilde{V}_\mu$, we can't reconstruct a unique $V^\mu$. Similarly, we can create an up indexed vector $\tilde{V}^\mu = V_{\nu} h^{\mu\nu}$. By definition of $\tilde{V}_\mu$ and $\tilde{V}^\mu$, it is evident that they will transform as (0,1) and (1,0) tensors respectively under any general coordinate transformation. Likewise, we can define an up indexed contravariant operator $\p^\mu$ from $\p_\mu$ by contracting it with $h^{\mu\nu}$ as
\be{p2}
\p^\mu = h^{\mu\nu} \p_\nu \quad \implies \p^\mu = (0,\p_x,\p_y,\p_z).
\ee
\subsection*{Covariant Formulation of GCA}
We will first review the Galilean Conformal Algebra (GCA) and its infinite extensions in component form and then write it down in covariant language. We can find the Galilean group by doing an Inonu-Wigner contraction \cite{Bagchi:2009my} on the conformal group. The process of going to the Galilean framework from relativistic one involves breaking of Lorentz covariance.
In order to do so, we will consider  $ \frac{x_i}{ct}\equiv \frac{\e x_i}{t} \rightarrow 0$.

The finite part of GCA (f-GCA) consists of rotations ($J_{ij}$), spacetime translations ($H$ and $P_i$), boosts ($B_i$), scaling ($D$) and special conformal transformations ($K$ and $K_i$). In terms of vector field, we can write the generators of the Galilean group as
\be{}\label{gp}
J_{ij} = - (x_i \p_j - x_j \p_i), \qquad
P_i = \p_i, \qquad H = -\p_t, \qquad
B_i = t\p_i.
\ee
The extension to a conformal part of the algebra is done through scaling ($D$) and special conformal transformations ($K$ and $K_i$). They are given by
\be{}\label{cp}
D = - (t\p_t + x^i\p_i), \qquad
K = - (2tx^i\p_i + t^2\p_t), \qquad K_i = t^2\p_i.
\ee
We can rewrite the generator \eqref{gp} and \eqref{cp} in a more compact form given by
\be{}\label{fs}
L^{(n)} = -(n+1)t^n x^i \p_i - t^{n+1}\p_t , \qquad M_i^{(n)} = t^{n+1} \p_i ,
\ee
where for $n=0,\pm 1$, the generators $L^{(n)},M^{(n)}_{i}$ denotes
\bea{}\label{f3}
L^{(-1,0,1)} = \{H, D, K\},~~ M_i^{(-1,0,1)} = \{P_i , B_i , K_i\} .
\eea
The f-GCA in terms of these new generators can now be written as following
\bea{}\label{f4}
&&\lb L^{(n)} , L^{(m)} \rb = (n-m)L^{(n+m)} ,~\lb L^{(n)}, M_i^{(m)} \rb = (n-m)M_i^{(n+m)} ,\non\\&&
\lb M_k^{(n)}, J_{ij} \rb = (M_i^{(n)} \delta_{jk} - M_j^{(n)} \delta_{ik}),~
\lb M_i^{(n)} , M_j^{(m)} \rb = 0,~\lb L^{(n)}, J_{ij} \rb = 0.
\eea
One can see that \eqref{f4} closes even if we let the index $n$ of \eqref{fs} run over all integers. It means that finite algebra becomes an infinite-dimensional algebra. We will refer to this algebra \eqref{f4} as GCA from now on. Another thing to notice from \eqref{fs} is that the generators $M_i^{(n)}$ give rise to time-dependent boosts, whereas $L^{(n)}$ generates some form of conformal isometry of the Galilean spacetime.
Similar to $L^{(n)}$ and $M^{(n)}_i$, the rotation generators could also be given an infinite lift as follows
\be{}
J_{ij}^{(n)} = -t^n (x_i\p_j - x_j \p_i).
\ee
The complete infinite-dimensional algebra becomes
\bea{}\label{f6}
&&\lb L^{(n)} , L^{(m)} \rb = (n-m)L^{(n+m)},~ \lb L^{(n)}, M_i^{(m)} \rb = (n-m)M_i^{(n+m)},\lb M_i^{(n)} , M_j^{(m)} \rb = 0,\non\\\non
&& \lb L^{(n)}, J_{ij}^{(m)} \rb = -m J_{ij}^{(n+m)},~
\lb J_{ij}^{(n)}, M_r^{(m)} \rb = -(M_i^{(n+m)} \delta_{jr} - M_j^{(n+m)} \delta_{ir}), \non\\&&
\lb J_{ij}^{(n)} , J_{rs}^{(m)} \rb = \delta_{is} J_{rj}^{(n+m)} + \delta_{jr} J_{si}^{(n+m)} + \delta_{ir} J_{js}^{(n+m)} + \delta_{js} J_{ir}^{(n+m)}.
\eea
\\
We will now use the results from above to write the generators \eqref{fs} and the algebra \eqref{f4} in a covariant form.
We have seen that the variable $t$ appears in most of the generators of GCA. Our first step will be to write $t$ in a covariant tensor form. We also know that time is absolute in Galilean relativity and doesn't change under time-independent Galilean boosts and rotation. Therefore, $t$ is a scalar and given by
\bea{}
t = x^\a \t_\a .
\eea
We will note down the following trivial, nevertheless, useful properties of $t$:
\bea{}
\p_\mu t = \t_\mu,~
\p_\mu t^n = n t^{n-1} \t_\mu,~
\p^\mu t^n = 0 .
\eea
Now, we will move on to defining the generators in covariant formulation. Let us first focus on time-dependent rotation and boosts. They are given by
\bea{}
J^{ij (n)} = -t^n (x^i \p^j - x^j \p^i),~
J^{0 j \; (n)} = -t^{n+1} \p_j = - M_j^{(n)}.
\eea
Both of the generators can be combined into a single generator as
\bea{}
J^{\mu\nu\;(n)} = -t^n (x^\mu \p^\nu - x^\nu \p^\mu).
\eea
Now will write down $L^{(n)}$ in covariant form. For that, we will examine its transformation under a Galilean boost $(x^i \rightarrow x^i-v^i t)$ given by
\bea{}
L^{(n)} \rightarrow L^{(n)} + v^i (n t^{n+1} \p_i).
\eea
It looks like a transformation of the zeroth component of a down indexed Galilean vector. Let us denoted it by $Z_{\mu}$. Then the transformation looks like
\be{}
 Z_0 \rightarrow Z_0 + v^i Z_i .
\ee
Inspired by this, we define a down indexed vector $Z_{\mu}$ as follows
\be{}
Z_\mu^{(n)} = -(n+1)t^n \t_\mu (x^\a \p_\a) + n t^{n+1} \p_\mu .
\ee
It is easy to verify that $Z_0^{(n)} = L^{(n)}$ and $Z_i^{(n)} = n M_i^{(n)}$.
Note that these two generators $Z^{(n)}_\mu$ and $J^{\mu\nu\;(n)}$ are not independent, but are related to the time dependent boosts $M_i^{(n)}$ as
\bea{}
Z_i^{(n)} = n J^{i0\;(n)} = n M_i^{(n)}.
\eea
In covariant form, this dependence is given by
\bea{}
h^{\mu\nu}Z_\mu^{(n)} = -n \; \t_\mu J^{\mu\nu\;(n)}.
\eea
Next, we can write down the commutation relations of these generators. They are given by 
\bea{}\label{f7}
&&\lb J^{\mu\nu\;(n)} , J^{\a\b\;(m)} \rb = h^{\mu \a} J^{\nu\b\;(n+m)} + h^{\mu \b} J^{\a\nu\;(n+m)} +h^{\nu \a} J^{\b\mu\;(n+m)} + h^{\nu \b} J^{\mu \a\;(n+m)},\non\\\non
&&\lb Z_{\mu}^{(n)}, Z_{\nu}^{(m)} \rb = \frac{n-m}{2} \left( \t_\nu Z_\mu^{(n+m)} +\t_\mu Z_\nu^{(n+m)} \right) + \frac{(n-m)^2}{2}T_{\mu\a\b\nu} J^{\a\b \; {(n+m)}}, \\
&&\lb Z_{\mu}^{(n)} ,J^{\a\b\;(m)} \rb = -m \t_\mu J^{\a\b\;(n+m)} + n \; \t_\gamma \left(J^{\gamma \b \;(n+m)} \delta^\a_\mu - J^{\gamma \a \; (n+m)} \delta^\b_\mu \right).
\eea
where $T_{\a\b\mu\nu}$ is defined in \refb{T}.
We can also split the commutation $[Z_{\mu}^{(n)}, Z_{\nu}^{(m)}]$ for the two cases when $m +n \neq 0$ and $m + n = 0$ as
\bes \label{}
\bea{}
&&\lb Z_{\mu}^{(n)}, Z_{\nu}^{(m)} \rb = \frac{n-m}{n+m}\left( n \t_\nu Z_\mu^{(m+n)} + m \t_\mu Z_\nu^{(m+n)} \right) \qquad \text{for} \;\; m \neq -n\\
&&\lb Z_{\mu}^{(n)}, Z_{\nu}^{(-n)} \rb = 2n \left(\t_\mu Z_\nu^{(0)} + n\;T_{\mu\a\b\nu} {J}^{\a\b\;(0)} \right) \qquad\qquad \text{for} \;\; m = -n
\eea
\ees
If we restrict the rotation $J^{\mu \nu\;(n)}$ only to time-independent rotations i.e. restrict $n=0$ and define $J^{\mu\nu}:= J^{\mu \nu \; (0)}$, then we get an infinite-dimensional subalgebra as following:
\bea{} \label{f8}
&&\lb Z_{\mu}^{(n)}, Z_{\nu}^{(m)} \rb = \frac{n-m}{n+m}\left( n \t_\nu Z_\mu^{(m+n)} + m \t_\mu Z_\nu^{(m+n)} \right) \qquad \text{for} \;\; m \neq -n\non\\\non
&&\lb Z_{\mu}^{(n)}, Z_{\nu}^{(-n)} \rb = 2n \left(\t_\mu Z_\nu^{(0)} + n\;T_{\mu\a\b\nu} {J}^{\a\b} \right), \\\non
&&\lb Z_{\mu}^{(n)} ,J^{\a\b} \rb = \; \left(h^{\gamma \a} \delta^\b_\mu - h^{\gamma \b} \delta^\a_\mu \right) Z_{\gamma}^{(n)},\\
&&\lb J^{\mu\nu} , J^{\a\b} \rb = h^{\mu \a} J^{\nu\b} + h^{\mu \b} J^{\a\nu} +h^{\nu \a} J^{\b\mu} + h^{\nu \b} J^{\mu \a}.
\eea
In the subsequent sections, we will show that this infinite-dimensional subalgebra will come out to be the symmetry of Galilean electrodynamics at the level of Lagrangian.

\subsection*{Representation of GCA for Galilean Vectors}\label{repv}
We will now write down the representation of GCA for Galilean vectors. The highest weight representation was discussed at lengths in \cite{Bagchi:2009ca, Bagchi:2014ysa, Bagchi:2015qcw, Bagchi:2017yvj}. Here, we will construct it in the language of covariant formalism. The theory of representation will then serve us to determine the invariance of the Lagrangian under GCA. In terms of notations, we will use $\lb T, \Phi(t,x) \rb$ for the action of a group generator $T$ on a generic field $\Phi(t,x)$. Since we will be talking about gauge fields ($V^{\mu}, V_{\mu} \equiv a^{\mu}, a_{\mu}$) in this paper, we will only note down the representation theory in terms of these fields.

Let us start by writing down the action of $J^{\mu\nu\,(0)}$ (or simply $J^{\mu\nu}$) on vectors $a^\a$ and $a_\a$. It comes out to be
\bea{}&&\label{jtrans}
\lb J^{\mu\nu} , a^\a \rb = (x^\mu \p^\nu - x^\nu \p^\mu) a^\a + (\delta^{\nu}_\b \; h^{\a \mu} - \delta^{\mu}_\b \; h^{\a\nu}) a^\b,\non\\&&
\lb J^{\mu\nu} , a_\a \rb = (x^\mu \p^\nu - x^\nu \p^\mu) a_\a + (\delta^{\mu}_\a \; h^{\b \nu} - \delta^{\nu}_\a \; h^{\b\mu})a_\b.
\eea
If you write \eqref{jtrans} in component form, you will get the same expressions mentioned in \cite{Bagchi:2014ysa}. Similarly, the action of $Z_\mu^{(n)}$ is given by
\bes{}\label{zsym}
\bea{7.30}&&
\lb Z_\mu^{(n)} , a^\a \rb = \Big((n+1)t^n \t_\mu (x^\b \p_\b) - nt^{n+1}\p_\mu + (n+1)t^n \t_\mu\Big)a^\a \non\\&&\hspace{4.2cm}+n(n+1)t^{n-1} a^\b x^{\gamma} \Big( \delta^\a_\gamma \tau_{\b\mu} - \delta^\a_\mu \t_{\b\gamma}\Big),\\\non&&
\label{7.31}
\lb Z_\mu^{(n)} , a_\a \rb = \Big((n+1)t^n \t_\mu (x^\b \p_\b) - nt^{n+1}\p_\mu + (n+1)t^n \t_\mu\Big)a_\a \non\\&&\hspace{4.2cm}+ n(n+1)t^{n-1} a_\b x^{\gamma} \Big( \delta^\b_\mu \tau_{\gamma\a} - \delta^\b_\gamma \t_{\mu\a}\Big).
\eea\ees
For simplicity, we can also write \eqref{zsym} as follow
\bes
\bea{7.32}&&
\lb Z_\mu^{(n)} , a^\a \rb = (n+1)\t_\mu \left(t^n x^\b \p_\b a^\a + t^n a^\a - n (a^\b \t_\b) x^\a t^{n-1} \right) \non\\&&\hspace{4.05cm}- n\left(t^{n+1}\p_\mu a^\a - (n+1)t^n(a^\b \t_\b)\delta^\a_\mu \right),\\&&
\lb Z_\mu^{(n)} , a_\a \rb = (n+1)\t_\mu \left(t^n x^\b \p_\b a_\a + t^n a_\a + n (a_\b x^\b)\t_\a t^{n-1} \right) \non\\&&\hspace{4.65cm}- n\Big(t^{n+1}\p_\mu a_\a + (n+1)t^n a_\mu \t_\a \Big).
\eea\ees
where we have used \eqref{15} in the intermediate steps. Similar to $J^{\mu\nu}$, if we replace $Z_\mu^{(n)}$ with $(L^{(n)},n M^{(n)}_{i})$ and write the gauge fields in component forms, we get back the expressions which are mentioned in \cite{Bagchi:2014ysa}.
\section{Galilean Electrodynamics: From perspective of equations of motion}
In literature, Galilean electrodynamics (GED) was studied first by Le Bellac and Levy- Leblond in \cite{LBLL}. In \cite{Bagchi:2014ysa}, the authors were motivated to search for finite and infinite GCA symmetries in GED at the level of equations of motion. To their surprise, they find GCA as symmetries of this theory.

In this section, we will construct GED in the covariant formulation. We will first start by briefing ourselves on the equations of relativistic electrodynamics (Maxwell's equations) that are given by
\be{26}
{\p_r}_\nu F_r^{\mu\nu} = \mu_0 J_r^\mu ,
\ee
where $F_r^{\mu\nu}=(\p_r^{\mu}A_r^{\nu}-\p_r^{\nu}A_r^{\mu})$ is the electromagnetic tensor, $J_r^{\mu}$ is the four-current, ${\p_r}_\nu = (\p_t/c , \p_i)$ is the relativistic partial derivative. Next, depending on whether $A_r^\mu$ and $J_r^\mu$ are largely timelike or largely spacelike, we will redefine $A_r^\mu$ and $J_r^\mu$ in terms of new fields $a^\mu$ or $a_\mu$ and $j^\mu$ or $j_\mu$ respectively which will act as the Galilean version of $A_r^\mu$ and $J_r^\mu$.

To begin with, we will unpack \eqref{26} in the space and time components and write $c=1/\e$:
\bea{}
-\epsilon\p_t (\p_i A_r^i)-\p_i\p^i A_r^0 = \mu_0 J_r^0,\non\\\label{28}
\p^j (\epsilon \p_t A_r^0+\p_i A_r^i)-(-\epsilon^2 \p_t + \p_i\p^i) A_r^j = \mu_0 J_r^j .
\eea
We will now invoke the two kinds of limits defined in  Sec[\ref{sdll}] on $A_r^\mu$ and $J_r^\mu$. The largely timelike limit on $A_r^\mu$ and $J_r^\mu$ is known as the Electric limit, whereas the largely spacelike limit on $A_r^\mu$ and $J_r^\mu$ is the Magnetic limit \cite{Bagchi:2014ysa}.

\subsection*{Electric Limit}
In terms of $\epsilon$, the largely timelike limit on the gauge field and the current in \refb{28} is the following
\bea{29}
A_r^0 = \frac{a^0}{\epsilon},~ A_r^i = a^i;\qquad J_r^0 = \frac{j^0}{\epsilon},~ J_r^i = j^i.
\eea
Plugging \eqref{29} into \eqref{28} and taking the limit $\epsilon \to 0$, the final resulting equations becomes
\bea{33}
-\p_i\p^i a^0 = \mu_0 j^0,~
\p^j (\p_t a^0+\p_i a^i)-(\p_i\p^i) a^j = \mu_0 j^j .
\eea
The equations \eqref{33} are the non-relativistic limit of Maxwell's equations in the Electric limit \cite{Bagchi:2014ysa}.

We will now write the equations in the language of Galilean vectors as follows. Let us define a quantity analogous to the electromagnetic tensor as:
\bea{35}
f^{\mu\nu} = (\p^\mu a^\nu - \p^\nu a^\mu),
\eea
where $\p^\mu = (0 , \p_x, \p_y, \p_z)$ as defined in \refb{p2} and $a^\mu$ is the Galilean definition of gauge field in this limit. In terms of $f^{\mu\nu}$, the equations \refb{33} becomes:
\be{36}
\p_{\nu} f^{\mu\nu} = \mu_0 j^\mu.
\ee
\subsection*{Magnetic Limit}
Like in the previous case, we need to invoke the spacelike limit on $(A_r^\mu,J_r^\mu)$ in \refb{28}. In terms of $\epsilon$, the spacelike condition becomes
\bea{37}
A_r^0 =- \epsilon a_0,~ A_r^i = a_i;\qquad
J_r^0 =- \epsilon j_0,~ J_r^i = j_i.
\eea
Plugging \eqref{37} into \eqref{28} and using the limit $\epsilon \to 0$, the result come out to be
\bea{41}
\p_t \p_i a_i-\p_i\p^i a_0 = \mu_0 j_0,~
\p^j \p_i a_i-\p_i\p^i a_j = \mu_0 j_j.
\eea
These are the equations of motion in the Magnetic limit. Similar to the Electric limit, we will define a quantity analogous to the electromagnetic tensor:
\bea{43}
f_{\mu\nu} = (\p_\mu a_\nu - \p_\nu a_\mu),
\eea
where $\p_\mu = (\p_t , \p_x, \p_y, \p_z)$ and $a_\mu$ is the Galilean definition of gauge field in the Magnetic limit.
In terms of this $f_{\mu\nu}$, the equations \refb{41} become very compact as:
\bea{44}
\p^{\nu} f_{\mu\nu} = \mu_0 j_\mu.
\eea
\subsection*{The General Solution of the GED}
We now have the equations of motion that we calculated in the previous section. The next step will be to find the general solutions in the presence of the sources for both the limits. To solve \eqref{36} and \eqref{44}, we will use a particular gauge choice, which is analogous to the Lorenz gauge (${\p_r}_\mu A^{\mu}_r=0$). In other words, we will choose a gauge that will hold in all inertial frames. Specifically, we choose the following gauges
\bes{} \label{4.9}
\bea{}
\p_\mu a^\mu &= 0 \quad(\text{Electric limit}),\\
\p^\mu a_\mu &= 0 \quad(\text{Magnetic limit}),
\eea\ees
for our case. It is easy to show that the Lorenz gauge in the non-relativistic limit becomes one of the above two depending on the Electric or Magnetic limit. We first show that we can choose such gauge conditions with some assumptions. Suppose that we know a field configuration $a^\mu(x^\mu)$ (in the case of Electric limit) that doesn't obey the above gauge conditions. Because of the gauge freedom, let us see if we can add a gauge term $\p^\mu X$ for some function $X$ such that the new field obeys the above gauge conditions and so the following will hold
\bea{4.11}
\p_\mu (a^\mu + \p^\mu X) = 0 \implies
\p_\mu\p^\mu X =- \p_\mu a^\mu.
\eea
Note that $\p_\mu \p^\mu = \nabla^2$ is the Laplacian operator in Galilean definitions. Given the values of $a^\mu$, the above condition is a Poisson equation in the variable $X$. We assume that the solution for this $X$ exists which means, we can choose the gauge conditions as \refb{4.9}.

Now we solve for the non-relativistic equations \refb{36} and \refb{44} which on imposing the above gauge conditions \refb{4.9} becomes
\bea{4.13}
\p_\mu \p^\mu a^\nu = - \mu_0 j^\nu \quad \text{(Electric limit)},~
\p_\mu \p^\mu a_\nu = - \mu_0 j_\nu \quad \text{(Magnetic limit)}.
\eea
These are Poisson equations and can be solved to obtain the following solutions
\bes{}\label{4.15}
\bea{}&&
a^\nu(x^\a) = \mu_0 \int{\frac{j^\nu (x'^\a)}{r} d^4x'}\quad \text{(Electric limit)}, \\&&
a_\nu(x^\a) = \mu_0 \int{\frac{j_\nu (x'^\a)}{r} d^4x'}\quad \text{(Magnetic limit)}.
\eea\ees
where $r$ is the spatial distance between the two simultaneous events ${x'}^\mu$ and $x^\mu$.
The events ${x'}^\mu$ and $x^\mu$ are simultaneous because there is no time derivative in the Poisson equation.
Hence, these solutions are in stark contrast with the relativistic solutions. Fields at one point are instantaneously affected by a source at some arbitrary distance. Because the difference of ${x'}^\mu$ and $x^\mu$ has time difference equal to zero, the vector ${x'}^\mu - {x}^\mu$ belongs to the space $T_0$ defined in \eqref{T0}. The distance $r$ is defined using the submetric tensor $h_{\mu\nu}$ as
\be{4.17}
r^2 = (x'^\mu - x^\mu)h_{\mu \nu}(x'^\nu - x^\nu).
\ee
We will get the same results as we got in \refb{4.15} if one takes the non-relativistic limit on the solutions of Maxwell's equation written in Lorentz gauge. \section{Action of Galilean Electrodynamics}
The goal here is to write down the action in terms of covariant formulation in the Galilean background. To evaluate it, we will use the equations of motion of GED denoted by
\bea{sds}
\p_{\nu} f^{\mu\nu} = \mu_0 j^\mu,~\p^{\nu} f_{\mu\nu} = \mu_0 j_\mu.
\eea
Let us begin with the action of relativistic electrodynamics in $d=4$ given by
\be{} S_r ({A_r}_{\mu}, {\p_r}_{\mu}{A_r}_{\nu})=\int d^4x\, \Big[-\frac{1}{4}{F_r}_{\mu\nu}{F_r}^{\mu\nu} +\mu_0 {J_r}^{\mu} {A_r}_{\mu}\Big],
\ee
which gives the equations of motion \eqref{26} where the subscript $r$ denotes relativistic definition of quantities. Similarly, we want to find the action which gives \eqref{sds} as the equations of motion in Galilean limit.
The action which gives us the correct non-relativistic equations is
\bea{4.2}\label{lagnr} S(a_{\mu}, a^{\mu}, \p_{\mu}a_{\nu}, \p^{\mu}a^{\nu})= \int d^3x\,dt\,\Big[-\frac{1}{4}f_{\mu\nu}f^{\mu\nu} +\frac{1}{2}\mu_{0}\,j^{\mu} a_{\mu}+\frac{1}{2}\mu_0\, j_{\mu}a^{\mu}\Big].
\eea
Notice that we have written the action as a function of $a^{\mu}$ and $a_{\mu}$. It means that $a^{\mu}$ and $a_{\mu}$ are two independent fields, and there is no invertible linear map between them. To find the equations of motion from \eqref{lagnr}, we have to treat the two fields separately. Mathematically speaking, the following functional derivatives vanish:
\bea{}\frac{\delta a^{\mu}}{\delta a_{\nu}}=0 ~\text{and}~ \frac{\delta a_{\nu}}{\delta a^{\mu}}=0.\eea
It is easy to see from the Lagrangian \eqref{lagnr} that we get the equations \eqref{sds} for the Electric and the Magnetic limit by varying the action with respect to $a_{\mu}$ and $a^{\mu}$ respectively.
\subsection*{Gauge Transformation}
From this section onwards, we will only consider the free Lagrangian (i.e. without current sources $j^\mu$ and $j_\mu$) unless mentioned. The reason is that the source term can be different matter (scalars or fermions) fields and these fields transform differently under symmetry transformations. To avoid the difficulties related to it right now, we will stick only to the free part of the Lagrangian.

In the non-relativistic limit, two sets of gauge symmetries are also present. They are somehow similar to the relativistic case in appearance. It is because both $f_{\mu\nu}$ and $f^{\mu\nu}$ appear in the free Lagrangian as
\bea{fp}
\mathcal{L}_f = -\frac{1}{4} f^{\mu\nu}f_{\mu\nu}.
\eea
The two sets of gauge symmetries, one for the Electric case and one for the Magnetic case, are as follows:
\bea{}
&& a^\mu \to a^\mu + \p^\mu X \implies f^{\mu\nu} \to f^{\mu\nu}\quad \text{(Electric Limit)},\label{elgi}\\
&& a_\mu \to a_\mu + \p_\mu Y \implies f_{\mu\nu} \to f_{\mu\nu} \quad \text{(Magnetic Limit)}\label{mlgi}.
\eea
Because these are two independent sets of symmetries, we have used two different variables that are $X$ and $Y$ for the gauge transformations. Also, in the Electric case, $a^\mu$ is transformed by $\p^\mu X = (0, \p^1 X, \p^2 X, \p^3 X)$ which means only the three spatial components are transforming, unlike the Magnetic case.

\subsection*{Symmetries of the Lagrangian}\label{sol}
We will now move on to finding the symmetries related to the free Lagrangian \eqref{fp}. For that, we will use the results mentioned in Sec[\ref{repv}]. Because of the covariant formalism, it is evident that the Lagrangian of the free theory is invariant under Galilean boosts, rotations and translations. To find the variation of Lagrangian under $Z^{(n)}_\mu$, denoted by $\delta[Z^{(n)}_\mu] \mathcal{L}_f$, we need to use \refb{zsym} which describes the variation of the gauge fields $a_\mu$ and $a^\mu$ under $Z^{(n)}_\mu$ and a simple calculation shows:
\bea{}\label{Zc1}
&&\delta[Z^{(n)}_\mu]\mathcal{L}_f = (n+1)\t_\mu t^n \p_\a\left( x^\a \mathcal{L}_f \right) - n t^{n+1} \; \p_\mu \mathcal{L}_f \non\\&&\hspace{1.5cm}=
\p_\a \left( (n+1)\t_\mu t^n x^\a \mathcal{L}_f -\delta^\a_\mu \;n \; t^{n+1} \mathcal{L}_f \right).
\eea
In the second line, we have used the by-parts rule to get the variation as a total derivative term. In conclusion, it means that the generator $Z^{(n)}_\mu$ is a symmetry of the Lagrangian.
%
\section{Energy-Momentum Tensor} \label{EMT}
We will derive the energy-momentum tensor (EM tensor) for the Galilean electrodynamics using the similar procedures that we followed for finding the non-relativistic limit of Maxwell's equations. We start by writing the EM tensor of the relativistic theory given by
\bea{3.26}
{(T_r)}^\mu_\a = {(F_r)}^{\mu\nu}{(F_r)}_{\nu\a} + \frac{1}{4}\delta^\mu_\a \left({(F_r)}^{\b\nu}{(F_r)}_{\b\nu}\right),
\eea
and then take the non-relativistic limit on the relativistic EM tensor and write the resultant in terms of the Galilean vectors $a^\mu$ and $a_\mu$.

As we know that the EM tensor can be written in the relativistic scenario as $(T_r)^{\mu\nu}$, $(T_r)^\mu_\nu$ or $(T_r)_{\mu\nu}$. We can switch between the up and down indices using the Minkowski metric $\eta_{\mu\nu}$ or $\eta^{\mu\nu}$. Unlike the relativistic case, our Galilean scenario doesn't have the freedom to switch between such tensors. It is because the metric tensors are non-invertible and so any contraction of some tensor with the metric tensor will lose information of that tensor. So, we can only have one unique definition of EM tensor, which will have the full information of the energy-momentum of the fields, instead of three inter-convertible tensors ($T^{\mu\nu}$, $T^\mu_\nu$ or $T_{\mu\nu}$).

To find out which one of these three is mathematically appropriate, we propose the following arguments. The EM tensor is usually of interest partly because of its conservation law. Mathematically, we take a divergence of the EM tensor to find the rate of change of energy-momentum of the fields such that it is equal to zero for a free theory. In our Galilean framework, we can represent the divergence by a contraction with $\p_\mu$ which implies that the EM tensor must have an upper index. We cannot use $\p^\mu = (0 , \p_x, \p_y , \p_z)$ because there is no time-derivative in $\p^\mu$. So we have to take the contraction with $\p_\mu$ to be able to talk about the conservation law.

The divergence of the EM tensor does not vanish in the presence of the source. Mathematically, the divergence of the EM tensor is equal to the negative of the Lorentz force $(\mathcal{F}_{\nu})$ applied by the fields on the sources. It is given by
\bea{}\label{tuv}{\p_r}_\mu (T_r)^\mu_\nu = -(\mathcal{F}_r)_\nu.\eea
If we use this definition of the Lorentz force, we see that we have to perform a contraction on the EM tensor with $\p_\mu$. It is because of the force given by a down indexed vector in the Galilean case. It leaves us with only one thing that is $T^\mu_\nu$.
\subsubsection*{Electric Limit}
We have to substitute \eqref{29} in place of the field $A_r^\mu$ to write it in terms of the Galilean vector $a^\mu$. But instead of this, we can also substitute $A^\mu_r = (a^0 , \e a^i)$. It is just a scaled definition of \eqref{29} by a factor $\e$. We do this to avoid $\e$ in the denominators and doing this won't change the physical content of EM tensor as it is homogeneous in $A^\mu_r$ (scaling $A^\mu_r$ by any number $k$, scales the EM tensor by $k^2$ and therefore we can extract the same useful content if we didn't scale). We simplify the first term on the right side of equation \refb{3.26} which is
\bea{3.27}\small{
{(F_r)}^{\mu\nu}{(F_r)}_{\nu\a} = \begin{bmatrix*}[c]
0 & -(\p^j a^0 + \e^2\p_t a^j)\\
\p^i a^0 + \e^2\p_t a^i & \e(\p^i a^j - \p^j a^i)
\end{bmatrix*}
\begin{bmatrix*}[c]
0 & \p_k a^0 + \e^2\p_t a^k\\
-(\p_j a^0 + \e^2\p_t a^j) & \e(\p_j a^k - \p_k a^j)
\end{bmatrix*}}.\non
\eea
In each matrix element, we will keep only the lowest order terms of $\e$, because higher order terms will inevitably go to zero on taking the limit $\e \to 0$. So, neglecting the higher order terms in $\e$, we can simplify the above equation as
\bea{3.28}
\small{{(F_r)}^{\mu\nu}{(F_r)}_{\nu\a} = \begin{bmatrix*}[c]
0 & -\p^j a^0\\
\p^i a^0 & \e(\p^i a^j - \p^j a^i)
\end{bmatrix*}
\begin{bmatrix*}[c]
0 & \p_k a^0\\
-\p_j a^0 & \e(\p_j a^k - \p_k a^j)
\end{bmatrix*}}.
\eea
We can also write it as
\bea{3.29}
\small{{(F_r)}^{\mu\nu}{(F_r)}_{\nu\a} = \begin{bmatrix*}[c]
\p^j a^0 \p_j a^0 & \e \p^j a^0 (\p_j a^k - \p_k a^j) \\
-\e(\p^i a^j - \p^j a^i)\p_j a^0 & \p^i a^0\p_k a^0
\end{bmatrix*}}.
\eea
Using the same process, we can simplify the second term $\frac{1}{4}\delta^\mu_\a {(F_r)}^{\b\nu}{(F_r)}_{\b\nu}$. It becomes 
\bea{3.30}
\small{\frac{1}{4}\delta^\mu_\a {(F_r)}^{\b\nu}{(F_r)}_{\b\nu} = -\frac{1}{2} \begin{bmatrix*}[c]
\p^j a^0 \p_j a^0 & 0 \\
0 & \delta^i_k \; (\p^j a^0 \p_j a^0)
\end{bmatrix*}}.
\eea
Now that we have the expressions in terms of Galilean vectors $a^\mu$ and $\e$. We need to remove $\e$ by defining a new tensor that will act as the Galilean version of the EM tensor. We will now look at the first term of ${(T_r)}^\mu_\a$ i.e. ${(F_r)}^{\mu\nu}{(F_r)}_{\nu\a}$. As described in Sec[\ref{GTs}], there is no direct way to extract a Galilean tensor out of it. To understand how to solve this problem, we need to use the fact that we are trying to find a specific tensor which is the EM tensor. More importantly, we want to find a conservation law that equates the divergence of the EM tensor and the Lorentz force on the sources. Let us first take divergence of the first term of ${(T_r)}^\mu_\a$ which is ${(F_r)}^{\mu\nu}{(F_r)}_{\nu\a}$ in the relativistic case. It is given by
\bea{3.31}
 \begin{bmatrix*}[c]
\e \p_t	\left(\p^j a^0 \p_j a^0\right) -\e \p_i \left(f^{ij}\,\p_j a^0 \right) & \;\;\; \e^2 \p_t \left(\p^j a^0 (\p_j a^k - \p_k a^j)\right) + \p_i \left(\p^i a^0\p_k a^0\right)
\end{bmatrix*},
\eea
where $f^{ij}=(\p^i a^j - \p^j a^i)$. We note that in the second element of the row matrix above, there is this term $\e^2 \p_t \left(\p^j a^0 (\p_j a^k - \p_k a^j)\right)$ which has to be neglected in comparison to $\p_i \left(\p^i a^0\p_k a^0\right)$ because of the $\e^2$ sitting in the former term. What we can do to neglect this is to remove the second element of the first row in the matrix of \refb{3.29} which is $\e \p^j a^0 (\p_j a^k - \p_k a^j)$. This may seem ad-hoc at first, but as we will show now it really solves the problem entirely. What we did is that we removed a term earlier which was eventually going to be removed. Therefore, we will write only the relevant terms (the ones which will survive eventually) of the matrix in \refb{3.29} as
\be{3.33}
{(F_r)}^{\mu\nu}{(F_r)}_{\nu\a} = \begin{bmatrix*}[c]
\p^j a^0 \p_j a^0 & 0 \\
-\e\,f^{ij}\p_j a^0 &\;\; \p^i a^0\p_k a^0
\end{bmatrix*}.
\ee
Now this looks like a quantity which can be converted to its Galilean counterpart. But we can do even better and neglect the irrelevant quantities (those which were going to be removed eventually) at one more earlier step i.e. in \refb{3.28}. In the equation \refb{3.28}, if we remove the second element of the second row of the second matrix i.e. $\e(\p_j a^k - \p_k a^j)$ and carry on the same steps forward, we get the same result as \refb{3.33}. Therefore, the relevant terms of \refb{3.28} are as follows:
\be{3.34}
 {(F_r)}^{\mu\nu}{(F_r)}_{\nu\a} = \begin{bmatrix*}[c]
0 & -\p^j a^0\\
\p^i a^0 & \e\,f^{ij}
\end{bmatrix*}
\begin{bmatrix*}[c]
0 & \p_k a^0\\
-\p_j a^0 & 0
\end{bmatrix*}.
\ee
Now it is relatively easier to guess the Galilean counterpart from here. We write the extracted Galilean parts from the useful components of ${T_r}^\mu_\a$ to formulate the Galilean EM tensor as
\be{EEM}
{T_E}^\mu_{\;\;\a} :=  f^{\mu\nu}\tilde{f}_{\nu\a} + \frac{1}{4} \delta^\mu_\a f^{\b\nu}\tilde{f}_{\b\nu},
\ee
where we define 
\bea{}
\tilde{f}_{\mu\nu} :=- f^{\a\b} T_{\mu\a\b\nu} = (\p_\mu \tilde{a}_\nu - \p_\nu \tilde{a}_\mu),~
\tilde{a}_\mu := a^\nu \t_{\mu\nu},
\eea
and the subscript $E$ denotes EM tensor in the Electric limit. It can be verified that this EM tensor is indeed conserved in the free theory and is also traceless. It is easy to show that in the presence of sources, the conservation equation comes out to be
\be{3.38}
\p_\mu {T_E}^\mu_{\;\;\a} = - j^\nu \tilde{f}_{\nu\a}.
\ee
where we have set $\mu_0=1$.
The definition of Lorentz force on the sources then becomes
$\mathcal{F}_\a:=-\p_\mu {T_E}^\mu_{\;\;\a}$
which in the Electric limit turns out to be
$\mathcal{F}_\a = j^\nu \tilde{f}_{\nu\a}$.
\subsubsection*{Magnetic Limit}
Using the same method as we did for the Electric limit, we will now write the EM tensor for the magnetic limit. 
After neglecting the irrelevant quantities in the case of magnetic limit, the first and the second term of the energy-momentum tensor come out to be the following
\bes{}\label{3.48}
\bea{}&&\hspace{-1cm}
\small{{(F_r)}^{\mu\nu}{(F_r)}_{\nu\a} = \begin{bmatrix*}[c]
	0 & 0\\
	-\e(\p_t a_j - \p_j a_0)f_{ij}& \;\;\;f_{ik}f_{jk}
\end{bmatrix*}},
\\&&\hspace{-1cm}
\small{\frac{1}{4}\delta^\mu_\a {(F_r)}^{\b\nu}{(F_r)}_{\b\nu} = -\frac{1}{2} \begin{bmatrix*}[c]
	(\p_j a_l - \p_l a_j) (\p_j a_l - \p_l a_j) & 0 \\
	0 &  \delta^i_k \;  (\p_j a_l - \p_l a_j) (\p_j a_l - \p_l a_j)
\end{bmatrix*}}.
\eea\ees
Using \eqref{3.48}, the Galilean EM tensor becomes
\be{MEM}
{T_M}^\mu_{\;\;\a} :=  \tilde{f}^{\mu\nu}f_{\nu\a} + \frac{1}{4} \delta^\mu_\a \tilde{f}^{\b\nu}f_{\b\nu},
\ee
where we define $
\tilde{f}^{\mu\nu} := h^{\a\mu}f_{\a\b}h^{\b\nu}$
and the subscript $M$ denotes EM tensor in the Magnetic limit. This tensor is conserved in the free theory and is also traceless. In the presence of sources, the conservation equation is given by
\be{}
\p_\mu {T_M}^\mu_{\;\;\a} = - \tilde{j}^\nu f_{\nu\a},
\ee
where we define $\tilde{j}^\nu := h^{\mu\nu}j_\mu$ and we have set $\mu_0 = 1$. The Lorentz force on the sources can be defined as
$\mathcal{F}_\a:=-\p_\mu {T_M}^\mu_{\;\;\a}$
which in the Magnetic limit turns out to be
$\mathcal{F}_\a = \tilde{j}^\nu f_{\nu\a}$.
\section{Noether Currents}
We have looked into the symmetries of the Lagrangian in Sec[\ref{sol}]. The next step will be to find the Noether charges corresponding to those symmetries.
For that, we will again look at the free Lagrangian \eqref{fp}.
We see that the Lagrangian contains both the Electric and the Magnetic limit. As we know, these two limits can't exist simultaneously by their definition. Still, if we find the Noether charges corresponding to a symmetry using the Lagrangian, we will get a conserved quantity containing both the $a^\mu$ and $a_\mu$, in general. At first sight, this doesn't make sense because any conserved quantity should refer to one physical scenario, either the Electric limit or the Magnetic limit. To understand the resolution of this seeming problem, we will first describe two properties to the non-relativistic equations.

\subsection*{Useful properties of GED}
To start, we will first write the properties in the case of Electric limit. We define two vectors $\tilde{a}_\mu$ and $\tilde{j}_\mu$. They are given by
\be{Adowntilde}
\tilde{a}_{\mu} = \t_{\mu\nu} a^\nu,~
\tilde{j}_{\mu} = \t_{\mu\nu} j^\nu.
\ee
In the previous section, we have already defined a tensor $\tilde{f}_{\mu\nu}$ while finding the EM tensor for the Electric limit as 
\be{Fdowntilde}
\tilde{f}_{\mu\nu} := -f^{\a\b} T_{\mu\a\b\nu} = (\p_\mu \tilde{a}_\nu - \p_\nu \tilde{a}_\mu),
\ee
in terms of the vectors \eqref{Adowntilde}. This $\tilde{f}_{\mu\nu}$ doesn't contain the full information of $f^{\mu\nu}$ by its definition. One can check that it still follow the equations of the Magnetic limit 
\be{C1}
\p^\nu \tilde{f}_{\mu\nu} = \mu_0 \tilde{j}_\mu.
\ee
We can prove it by multiplying $\t_{\a\nu}$ on both sides of equation of motion of the Electric limit. It shows that given a field configuration in the Electric limit, we can create a corresponding field configuration for the Magnetic limit by the procedure described above. This correspondence is not one to one as the information from converting $f^{\mu\nu}$ to $\tilde{f}_{\mu\nu}$ is lost.

Now we will state a similar properties for the Magnetic limit. We define two vectors $\tilde{a}^\mu$ and $\tilde{j}^\mu$ as
\be{Auptilde}
\tilde{a}^{\mu} = h^{\mu\nu} a_\nu,~
\tilde{j}^{\mu} = h^{\mu\nu} j_\nu.
\ee
We have defined a tensor $\tilde{f}^{\mu\nu}$ in the previous section and it is given by 
\be{Fuptilde}
\tilde{f}^{\mu\nu} := h^{\mu\a}f_{\a\b}h^{\nu\b} =  (\p^\mu \tilde{a}^\nu - \p^\nu \tilde{a}^\mu).
\ee
This $\tilde{f}^{\mu\nu}$ does not contain the full information of $f_{\mu\nu}$ by definition, still it follow the equations of the Electric limit. It is given by
\be{C2}
\p_\nu \tilde{f}^{\mu\nu} = \mu_0 \tilde{j}^\mu.
\ee
We can easily prove it by multiplying $h^{\a\nu}$ on both sides of the equations of motion of the Magnetic limit. It shows that given a field configuration in the Magnetic limit, we can create a corresponding field configuration for the Electric limit. These two properties \refb{C1} and \refb{C2} will be useful in the next section.

\subsection*{EM Tensor as a Noether Current}\label{EMNoether}
We will rederive the EM tensor for both the limits, but this time we will use the Noether's theorem. Using the spacetime translation symmetry in our GED Lagrangian, we obtain the canonical EM tensor denoted by $\theta^\mu_\nu (a^\gamma, a_\gamma )$, which is a function of both the dynamical fields $a_\mu$ and $a^\mu$, as
\bea{}&&\label{emt}
\theta^\mu_\nu (a^\gamma, a_\gamma ) = \frac{\p \mathcal{L}_f}{\p_\mu a^\a} \p_\nu a^\a + \frac{\p \mathcal{L}_f}{\p_\mu a_\a} \p_\nu a_\a - \delta^\mu_\nu \mathcal{L}_f
\non\\&&\hspace{1.8cm}
= -\frac{1}{2}h^{\mu\b}f_{\b\a} \p_\nu a^\a - \frac{1}{2}f^{\mu\a} \p_\nu a_\a - \delta^\mu_\nu (-\frac{1}{4} f^{\a\b}f_{\a\b}).
\eea
We will now describe what quantity we can extract from \eqref{emt}. Let us start by saying that we want to find the EM tensor for the Electric limit. It is easy to verify explicitly that the divergence of this canonical EM tensor will vanish, provided we use the equations of motion of the Electric and Magnetic limit. When we are in the Electric limit, the field $a_\mu$ of Magnetic limit is like an auxiliary field with no physical significance. Similarly, when we are in the Magnetic limit, the field $a^\mu$ is a field with no physical significance. If we have some field configuration $a^\mu$ in the Electric limit, then any gauge field $a_\mu$ that obeys the equations of motion $\p^\mu f_{\nu\mu} = 0$ allows this canonical EM tensor to be conserved. Looking at the canonical EM tensor this way for the Electric limit makes it richer, in that we can have infinite possible conserved currents, that is one conserved current for each field configuration $a_\mu$ satisfying $\p^\mu f_{\nu\mu} = 0$. We can pick any one of those currents at our convenience. But to find the EM tensor for the Electric limit, we must use the information from only the gauge field $a^\mu$.

To resolve this issue, we use those properties that we presented in the previous section. We know from \refb{C1} that $\tilde{a}_\mu$ is a vector which satisfies the
equations for the Magnetic limit and more importantly $\tilde{a}_\mu$ is a vector that can be constructed entirely from $a^\mu$ itself. Therefore, we can insert $\tilde{a}_\mu$ in \eqref{emt} to obtain a conserved tensor which is in principle defined only using the field configuration $a^\mu$. The EM tensor \eqref{emt} becomes
\bea{}&&\label{emtel}
\theta^\mu_\nu (a^\gamma, \tilde{a}_\gamma) = -\frac{1}{2}h^{\mu\b} (\p_\a \tilde{a}_\b - \p_\b \tilde{a}_\a ) \p_\nu a^\a - \frac{1}{2}f^{\mu\a} \p_\nu \tilde{a}_\a - \delta^\mu_\nu \left(-\frac{1}{4} f^{\a\b}(\p_\a \tilde{a}_\b - \p_\b \tilde{a}_\a ) \right)\non\\&&
\hspace{1.8cm}= - f^{\mu\a} \p_\nu \tilde{a}_\a +  \frac{1}{4} \delta^\mu_\nu f^{\a\b} \tilde{f}_{\a\b}.
\eea
The EM tensor is conserved, gauge independent and is made only from the information of the field $a^\mu$. The equation \eqref{emtel} can also be written down in terms of the variables $f^{\mu\nu}$ and $\tilde{f}_{\mu\nu}$. To do this, we will add a conserved current $f^{\mu\a}\p_\a \tilde{a}_\nu$ (divergenceless with respect to index $\mu$) to \eqref{emtel}. The new EM tensor then becomes
\bea{TE}
{T_E}^\mu_\nu := \theta^\mu_\nu (a^\gamma, \tilde{a}_\gamma) + f^{\mu\a}\p_\a \tilde{a}_\nu = f^{\mu\a} \tilde{f}_{\a\nu} +\frac{1}{4} \delta^\mu_\nu f^{\a\b} \tilde{f}_{\a\b}.
\eea
As we see, it is the same traceless EM tensor defined in \refb{EEM}.

We will now follow a similar procedure for finding the EM tensor in the Magnetic limit. This time the field $a^\mu$ acts like an auxiliary field. Let us say we have a particular field configuration of $a_\mu$ on-shell, then for every field configuration $a^\mu$ which satisfies $\p_\mu f^{\nu\mu}=0$ we have a corresponding conserved current defined by the canonical EM tensor. From \refb{C2}, we know that such a field configuration is $\tilde{a}^\mu := h^{\mu\a}a_\a$ which is constructed entirely out of the Magnetic field variable $a_\mu$. Now, we replace $\tilde{a}^\mu$ with $a^{\mu}$ in \refb{emt} but we will also need to add a conserved tensor $\tilde{f}^{\mu\a} \p_\a a_\nu $ to make it gauge independent, then the EM tensor becomes
\bea{TM}
{T_M}^\mu_\nu := \theta^\mu_\nu(\tilde{a}^\gamma , a_\gamma) + \tilde{f}^{\mu\a} \p_\a a_\nu = \tilde{f}^{\mu\a}f_{\a\nu} + \frac{1}{4}\delta^\mu_\nu \tilde{f}^{\a\b}f_{\a\b}.
\eea
It is the same as the same traceless EM tensor defined in \refb{MEM} for the Magnetic limit.

\subsection*{Conserved Currents for the Conformal Symmetries}
We will first find the canonical conserved current associated to the $Z^{(n)}_\a$ symmetry which will be a function of both $a^\nu$ and $a_\nu$. We denote it by $J^\mu[Z^{(n)}_\a](a^\nu, a_\nu)$. The canonical conserved current that we obtain from Noether's theorem using \refb{Zc1} is
\be{zc}
J^\mu[Z^{(n)}_\a](a^\nu, a_\nu) = \frac{\p \mathcal{L}_f}{\p (\p_\mu a_\nu)} \lb Z^{(n)}_\a , a_\nu \rb + \frac{\p \mathcal{L}_f}{\p (\p_\mu a^\nu)} \lb Z^{(n)}_\a , a^\nu \rb - ((n+1)\t_\a t^n x^\mu - \delta^\mu_\a n t^{n+1}) \mathcal{L}_f,
\ee
where $\mathcal{L}_f = -\frac{1}{4} F^{\mu\nu}F_{\mu\nu}$.
This current is conserved if both $a^\nu$ and $a_\nu$ satisfy their on-shell conditions i.e. \refb{36} and \refb{44}. We will follow on the similar lines of reasoning for finding the current for both the limits that we did while finding the EM tensor in the previous subsection. For the Electric limit, we have to replace $a_\mu$ by $\tilde{a}_\mu$ so that the conserved current contains only the information of the field $a^\mu$. It is given by
\bea{}&&
J^\mu[Z^{(n)}_\a](a^\nu, \tilde{a}_\nu) =-f^{\mu\nu} \left( (n+1)t^n \t_\a (x^\b \p_\b) - nt^{n+1}\p_\a + (n+1)t^n \t_\a \right)\tilde{a}_\nu \non\\&&\hspace{6.5cm}+ \frac{\;t^n}{4} \Big((n+1)\t_\a x^\mu - \delta^\mu_\a n t \Big)f^{\nu\b}\tilde{f}_{\nu\b}.
\eea
We can also add a conserved quantity to $J^\mu[Z^{(n)}_\a](a^\nu, \tilde{a}_\nu)$ which is defined as
\be{}
{C_E}^\mu_\a := f^{\mu\nu}\p_\nu \left((n+1)t^n \t_\a x^\b \tilde{a}_\b - n t^{n+1} \tilde{a}_\a \right)
\ee 
so that we can write the resultant in terms of $f^{\mu\nu}$ and $\tilde{f}_{\mu\nu}$ as
\bea{}&&
{J_E}^\mu[Z^{(n)}_\a]:=J^\mu[Z^{(n)}_\a](a^\nu, \tilde{a}_\nu) + {C_E}^\mu_\a=\left( (n+1)t^n \t_\a x^\b  - nt^{n+1}\delta^\b_\a \right)f^{\mu\nu}\tilde{f}_{\nu\b} \non\\&&\hspace{6.5cm}+ \frac{\;t^n}{4} \Big((n+1)\t_\a x^\mu - \delta^\mu_\a n t \Big)f^{\nu\b}\tilde{f}_{\nu\b}.
\eea
The final expression becomes
\bea{}
{J_E}^\mu[Z^{(n)}_\a] = \((n+1)t^n \t_\a x^\b - n t^{n+1}\delta^\b_\a \){T_E}^\mu_\b,
\eea
where ${T_E}^\mu_\b$ is the EM tensor for the Electric limit defined in \refb{EEM}. Substituting $n=-1$ above will give us the EM tensor ${T_E}^\mu_\a$ for the Electric limit.
\newline
For the Magnetic limit, we have to replace $a^\nu$ by $\tilde{a}^\nu$ in \refb{zc} to get
\bea{}&&
J^\mu[Z^{(n)}_\a](\tilde{a}^\nu, a_\nu) = -\tilde{f}^{\mu\nu} \left( (n+1)t^n \t_\a (x^\b \p_\b) - nt^{n+1}\p_\a + (n+1)t^n \t_\a \right)a_\nu \non\\&&\hspace{6.5cm} + \frac{\;t^n}{4} \Big((n+1)\t_\a x^\mu - \delta^\mu_\a n t \Big) \tilde{f}^{\nu\b} f_{\nu\b}.
\eea
This is a gauge dependent conserved current. We can add a conserved current to it to make it gauge independent. That conserved quantity is given as
\bea{} {C_M}^\mu_\a := \tilde{f}^{\mu\nu}\p_\nu \left((n+1)t^n \t_\a x^\b a_\b - n t^{n+1} a_\a \right).\eea 
After adding it to $J^\mu[Z^{(n)}_\a](\tilde{a}^\nu, a_\nu)$ and simplifying, we get the gauge independent conserved current for the Magnetic limit as
\be{}
J_M^\mu[Z^{(n)}_\a] = \((n+1)t^n \t_\a x^\b - n t^{n+1}\delta^\b_\a \){T_M}^\mu_\b,
\ee
where ${T_M}^\mu_\b$ is the EM tensor for the Magnetic limit defined in \refb{MEM}. If we substitute $n=-1$ above, we recover the EM tensor for the Magnetic limit.

\section{Conclusion}
\subsection*{A quick summary}
We will now summarise the main results mentioned in the paper. Our main aim was to construct the GCA and the Lagrangian for Galilean electrodynamics in the covariant formulation of Galilean spacetime.

We first looked into the basics of the non-relativistic limit and talked about the geometry of Galilean spacetime. We then discussed the properties of the metric tensor and the submetric tensor. We moved on to writing the GCA and its representation based on the newly constructed covariant formulation. Once we have that in place, we wrote down the most general Lagrangian for the Galilean electrodynamics. We also checked the symmetry and found that it comes out to be invariant under infinite-dimensional GCA. Finally, we calculated the energy-momentum tensor and the conserved Noether currents associated with this theory.
\subsection*{Future directions}
Some obvious generalisations follow from the current work. Below is a list of a few of them.
\begin{itemize}
\item \underline{Galilean gauge theories}: The equations of motion and its symmetries for these theories have already been seen in \cite{Bagchi:2015qcw, Bagchi:2017yvj}. We would like to have an action formulation for Galilean gauge theories in the covariant formulation constructed in this paper. We also want to look at the conserved charges and Poisson brackets associated with the charges. It will confirm the presence of GCA at the level of the charges.

\item \underline{Carrollian limit}: 
We like to build a similar covariant formulation for the Carrollian limit $(c\rightarrow 0)$ of electrodynamics. It was formulated first in \cite{Duval:2014uoa, Bagchi:2016bcd} in terms of equations of motion. Then the action was constructed using Helmholtz conditions in \cite{Banerjee:2020qjj}. This theory is very captivating with different sectors, both of which exhibit infinite-dimensional BMS symmetries. It will be interesting to examine it in the covariant formulation.

\item \underline{Supersymmetric Galilean theories}: 
We want to investigate the Galilean version of $\mathcal{N} = 4$ $SU(N)$ supersymmetric Yang-Mills (SYM). The hope is that even if the Galilean conformal symmetries survive only at the classical version of electrodynamics and Yang-Mills theories. The conformal symmetries would hold out the quantum lift in the supersymmetric generalisation. The expectation would be to obtain the super-GCA \cite{Sakaguchi:2009de, Bagchi:2009ke, deAzcarraga:2009ch, Lukierski:2011az} in the non-relativistic sector of $\mathcal{N} = 4$ SYM. These symmetries may indicate the presence of a new integrable sub-sector. It would differ from the usual integrable planar sector.
\end{itemize}
\section*{Acknowledgement}
It’s a pleasure to thank Arjun Bagchi, Rudranil Basu and Kinjal Banerjee for numerous discussions and inputs related to the draft.
AM would like to thank Alexander von Humboldt Foundation (AvH) for the Humboldt Research Fellowship.
\newpage
\appendix{}
\section{Construction of Galilean Tensors}\label{GTs}
We will use the tensor products of contravariant and covariant Galilean vectors to construct the Galilean tensors and infer their transformation properties. Specifically, under general coordinate transformations, a Galilean tensor of rank $(n,m)$ should transform as
\bea{p3}
T'^{\a_1 \a_2 ... \a_n}_{\b_1 \b_2 ... \b_m}(x') = \frac{\p x'^{\a_1}}{\p x^{\gamma_1}} \frac{\p x'^{\a_2}}{\p x^{\gamma_2}} ...\frac{\p x'^{\a_n}}{\p x^{\gamma_n}} \;
\frac{\p x^{\eta_1}}{\p x'^{\b_1}} \frac{\p x^{\eta_2}}{\p x'^{\b_2}}...\frac{\p x^{\eta_3}}{\p x'^{\b_3}}
T^{\gamma_1 \gamma_2 ... \gamma_n}_{\eta_1 \eta_2 ... \eta_m} (x).
\eea
We saw before that we cannot get the correct Galilean transformations by naively taking $c \to \infty$ limit on Lorentz vectors $(V^{\mu}, V_{\mu})$. We need to have certain conditions on Lorentz vectors (Sec[\ref{sdll}]). Similarly, when we try $c \to \infty$ on Lorentz tensors, they should have some specific forms to transform like Galilean tensors. Using the fact that we can decompose every tensor into a sum of outer products of vectors, we can then find the expression of Lorentz tensors explicitly written in terms of $c$ that can be converted into Galilean tensors in the limit $c \to \infty$. From now, we will use $1/\e$ instead of $c$ and will take $\e \to 0$.

Let us take an example to understand this. Consider a tensor ${T_r}^\mu_\a$ and say that it is given by
\bea{}
{T_r}^\mu_\a = \begin{bmatrix*}[c]
-\e A & \; B\\
-\e^2 C & \; \e D
\end{bmatrix*}.
\eea
This tensor can be thought of as linear combinations of outer product of largely timelike Lorentz vectors $P_r^\mu = (P^0 , \e P^i)$ and largely spacelike Lorentz vectors ${Q_r}_\a = (-\e Q_0 , Q_i)$. Mathematically, it follows as
\be{}\label{pq}
{T_r}^\mu_\a = \sum P_r^\mu {Q_r}_\a = \begin{bmatrix*}[c]
-\e A & \; B\\
-\e^2 C & \; \e D
\end{bmatrix*},
\ee
where the summation $\sum$ denotes that we are taking linear combinations of such dyads. Similar to what we did for vectors, we can define a new Galilean tensor $T^\mu_{\;\;\a} = \begin{bmatrix*}[c]
A & \; B\\
C & \; D
\end{bmatrix*}$, whose components are independent of $\e$. Now, because we see that ${T_r}^\mu_\a$ was made up of $P_r^\mu$ and $ {Q_r}_\a$, we know how ${T_r}^\mu_\a$ transforms under Lorentz boosts with velocities $v \ll c$. It means we can find out how $A$, $B$, $C$, and $D$ transforms. Thus, one can easily show that the transformation of the tensor $T^\mu_{\;\;\a}$ is the same as that of a (1,1) Galilean tensor under Galilean boosts. Also, it is important to note that we may need to divide or multiply a Lorentz tensor by an overall factor of $\e$ to see how it is made up of largely timelike or spacelike Lorentz vectors. For example, consider the following quantity
\be{11}
{M_r}^\mu_\a = \begin{bmatrix*}[c]
- A & \; B /\e\\
-\e C & \; D
\end{bmatrix*}.
\ee
Although it looks different from ${T_r}^\mu_\a$, it is just a scaled version of ${T_r}^\mu_\a$ by a factor $1/\e$. Hence, the Galilean tensor that we will extract from ${M_r}^\mu_\a$, will be the same as $T^\mu_{\;\;\a} $. In conclusion, the overall factors of $\e$ or $1/\e$ are irrelevant in deciding the extracted Galilean quantity.

Till now, we have only looked at a specific example where the relativistic quantity ${T_r}^\mu_\a$ is in the form of a linear combination of the outer product of the vectors \eqref{pq}. In other words, the dyads were all of the same ranks that allowed us to write consistent tensor indices on the Galilean tensor $T^\mu_{\;\;\a}$.

But in some cases, it may happen that a Lorentz tensor, when written in the lowest order of $\e$, cannot be decomposed into a linear combination of dyads of the same type (largely spacelike and largely timelike). It would then imply that the Lorentz tensor has no Galilean counterpart. For example, consider
\be{12a}
{S_r}^\mu_\a = \begin{bmatrix*}[c]
- A & \; \e B\\
-\e C & \; D
\end{bmatrix*}.
\ee
If we break this tensor into two parts, then we will see that it is a sum of two Galilean tensors of different ranks which would imply that the overall quantity has no corresponding Galilean tensor:
\be{12b}
{S_r}^\mu_\a = \begin{bmatrix*}[c]
- A & \; 0/\e\\
-\e C & \; D
\end{bmatrix*} + \begin{bmatrix*}[c]
0 & \; \e B\\
0/\e & \; 0
\end{bmatrix*}.
\ee
Let us first focus on the first matrix in the equation above. It looks like a particular case of \refb{11} with $B=0$, which means that the Galilean tensor that we can extract from it is of the kind $X^\mu_{\;\;\a}$ (some Galilean tensor with $\mu$ index up and $\a$ index down). Now, look at the second matrix in \refb{12b}. The Galilean quantity we can extract from this would be of the type $Y_\mu^{\;\;\a}$ (Galilean tensor with the $\mu$ index down and $\a$ index up). Therefore, the sum of these two cannot be a Galilean tensor. It means that all relativistic tensors need not have a corresponding Galilean tensor. 
\bibliographystyle{JHEP}
\bibliography{main}
\end{document}